\documentclass[fleqn,10pt]{wlscirep}

\usepackage[T1]{fontenc}
\usepackage{blindtext}
\usepackage{amsfonts}
\usepackage{amsmath}
\usepackage{microtype}
\usepackage{graphicx}
\usepackage{overpic}
\usepackage{booktabs}
\usepackage{tabularx}
\usepackage{algorithm}
\usepackage[numbers, compress]{natbib}
\usepackage[all]{hypcap}
\hypersetup{citecolor=blue}
\hypersetup{linkcolor=blue}
\hypersetup{urlcolor=cyan}
\usepackage[nameinlink,capitalize]{cleveref}

\makeatletter
\newcommand{\settitle}{\@maketitle}
\makeatother

\usepackage{grffile}

\newcommand{\be}{\begin{equation}}
\newcommand{\ee}{\end{equation}}
\newcommand{\bea}{\begin{eqnarray}}
\newcommand{\eea}{\end{eqnarray}}

\newcommand{\f}[2]{\frac{#1}{#2}}

\newcommand{\ccup}[1]{\left\{#1\right\}}

\newcommand{\Graph}{G}

\newcommand{\Nedge}{E}

\newcommand{\NetMass}{S}
\newcommand{\Inc}{B}
\newcommand{\Inode}{v}
\newcommand{\Inodetwo}{u}

\newcommand{\Flux}{F}
\newcommand{\Iedge}{e}
\newcommand{\Icomm}{i}
\newcommand{\Ncomm}{M}

\newcommand{\Cond}{\mu}
\newcommand{\Lap}{L}
\newcommand{\Press}{p}

\renewcommand{\ref}[1]{[\ref{#1}]}
\crefname{Methods}{Methods}{Methods}

\usepackage[caption=false]{subfig}

\usepackage[normalem]{ulem} 

\newcommand{\kl}{Kirchhoff's}

\begin{document}

\title{Multicommodity routing optimization for engineering networks}


\author[1,*]{Alessandro Lonardi}
\author[2]{Mario Putti}
\author[1]{Caterina De Bacco}
\affil[1]{Max Planck Institute for Intelligent Systems, Cyber Valley, T{\"u}bingen 72076, Germany}
\affil[2]{Department of Mathematics ``Tullio Levi-Civita'', University of Padua, Via Trieste 63, Padua, Italy}

\affil[*]{alessandro.lonardi@tuebingen.mpg.de}

\begin{abstract}
Optimizing passengers routes is crucial to design efficient transportation networks. Recent results show that optimal transport provides an efficient alternative to standard optimization methods. However, it is not yet clear if this formalism has empirical validity on engineering networks. We address this issue by considering different response functions---quantities determining the interaction between passengers---in the dynamics implementing the optimal transport formulation. Particularly, we couple passengers' fluxes by taking their sum or the sum of their squares. The first choice naturally reflects edges occupancy in transportation networks, however the second guarantees convergence to an optimal configuration of flows.  Both modeling choices are applied to the Paris metro. We measure the extent of traffic bottlenecks and infrastructure resilience to node removal, showing that the two settings are equivalent in the congested transport regime, but different in the branched one. In the latter, the two formulations differ on how fluxes are distributed, with one function favoring routes consolidation, thus potentially being prone to generate traffic overload. Additionally, we compare our method to Dijkstra's algorithm to show its capacity to efficiently recover shortest-path-like graphs. Finally, we observe that optimal transport networks lie in the Pareto front drawn by the energy dissipated by passengers, and the cost to build the infrastructure.
\end{abstract}

\maketitle

\section{Introduction}

Finding optimal flow configurations in transport networks is an important problem in many real-world applications.
While natural systems like river basins \cite{sinclair1996,rinaldo1992minimum,rinaldo1993self,sun1994minimum,konkol2021interplay}, leaf venations \cite{ronellenfitsch2016global,ronellenfitsch2019phenotypes,katifori2010damage,xia2007formation}, or
slime molds \cite{Tero439,tero2008flow,tero2006computation,tero2007mathematical,yamada2000intelligence,bonifaci2012physarum, bonifaci2013short, bonifaci2017revised} involve transport of one type of mass only, e.g. water, this may not be the case in several engineering systems. For instance, routing data packets in communication networks, or passengers in urban transportation networks, requires multicommodity approaches where mass of different types interacts in a shared infrastructure, contributing to minimize one unique cost.

Despite their practical significance, multicommodity algorithms based on optimization routines are burdened by high computational complexity, caused by the simultaneous assignment of multiple commodities. Therefore, practitioners often rely on heuristics and approximations that lead to suboptimal solutions \cite{salimifard2020multicommodity}.
Distributed approaches like message-passing algorithms have demonstrated encouraging results \cite{yeung2013networking,yeung2013physics,altarelli2015edge,de2014shortest,xu2021scalable,yeung2012competition}, but remain computationally costly in scenarios where there is a large number of origin-destination pairs to be routed, or when the network is not sparse.

A promising approach is that of optimal transport theory. Recent studies \cite{lonardi2021designing,bonifaci2021physarum} have shown that this theoretical formalism can be adapted to address multicommodity scenarios, generalizing well-established results for uni-commodity models \cite{facca2016towards,facca2019numerics, facca2020physarum,facca2020branching,baptista2020network,baptista2021image,baptista2021convergence}.
The works of Lonardi \emph{et al.} \cite{lonardi2021designing} and Bonifaci  \emph{et al.} \cite{bonifaci2021physarum} focus on a theoretical characterization of the problem, drawing a formal connection between optimal transport and an equivalent dynamical system that is formulated in terms of physical quantities like conductivities and fluxes.
While preliminary results on multilayer transportation networks \cite{adinoyi2021optimal} suggest an empirical validity of this choice, questions remain open about its applicability in settings involving the transport of passengers.

In this work, we address this concern by studying the behavior of optimal transport approaches for multicommodity routing on urban transportation networks, with an empirical analysis on the Paris metro network. Our goal is to evaluate how different  cost functions impact the distribution of passenger flows. In detail, we search for stationary solutions of a dynamics where edge capacities---conductivities---grow as an increasing function of the total amount of passengers traveling on the edges. We investigate numerically the cases where the dependence between conductivities and fluxes is either the sum of the passengers traveling on an edge (its $1$-norm), or the sum of their squares (its $2$-norm). The first choice is more intuitive, since counting the total number of users in a network is a natural metric to evaluate its occupancy. However, in the second case it is possible to prove that the companion gradient flow used in the numerical solver admits a unique stationary solution  \cite{lonardi2021designing,bonifaci2021physarum}.

We design several experiments to investigate the main properties of optimal network configurations in the two cases. First, we observe that the $2$-norm tends to dilute more substantially passengers on the network, avoiding heavily trafficked routes. Second, we compare our model with Dijkstra's algorithm \cite{dijkstra1959note}, a popular approach for shortest-path minimization. We find that our method is a robust and efficient alternative to reproduce shortest-path-like networks. Furthermore, we test resilience to infrastructural failures, i.e., node and edge removal. Results show that the geographical locations of stations  together with their degree, are decisive factors. Finally, we observe that optimal networks lie in the Pareto front drawn by two fundamental driving forces: the energy dissipated by passengers' flows and the network infrastructural cost.

\section{Results and Discussion}
\subsection{Multicommodity routing on networks}\label{sec:dyn}

We design a routing optimization problem on a network $\Graph(V,E)$, where $V$ and $E$ are the sets of nodes and edges, and each edge has length $\ell_\Iedge > 0$. The edges are given a conventional orientation stored in a signed incidence matrix, with elements $\Inc_{\Inode \Iedge} = \{ +1, -1,0 \}$ if $\Inode$ is the head, the tail, or neither of them for edge $\Iedge$, respectively.  We model transportation of $\Ncomm \geq 1$ commodities through the network, each identified by an index $\Icomm$. We use them to differentiate passengers entering the network from different stations ($i \in V$), so that multiple users sharing the same path catalyze traffic congestion. Suppose that a commodity  $i$ has a mass rate $\NetMass_\Inode^\Icomm$ flowing into node $\Inode$ and outflows $\NetMass^\Icomm_\Inodetwo$ $\forall \Inodetwo \neq \Inode$, with $\sum_\Inode \NetMass_\Inode^\Icomm = 0$, $\forall i \in V$, to ensure that the system is isolated.

\begin{figure}[htbp]
\centering
\includegraphics[width=0.5\columnwidth]{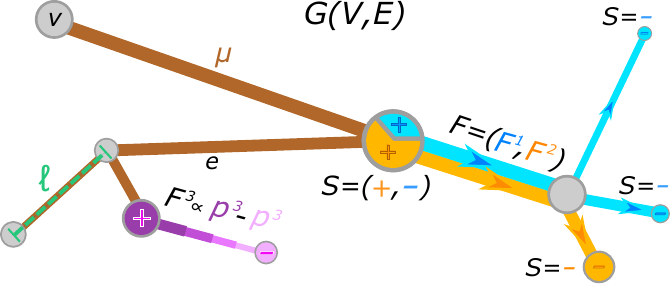}
\caption{\label{fig:model_representation} \textbf{Schematic problem visualization.} In brown we draw the edge capacities,  green is used to highlight the length of one edge. From the central node a positive inflow of two commodities enters (orange and light blue),  these move towards their destinations---the colored minuses---sharing an edge. Thus, multiple colored fluxes generate traffic congestion. In pink we represent the pressure potentials of a third commodity. Differences of pressure along an edge trigger $F^3$.}
\end{figure}

The main quantities of interest are the edge conductivities $\mu_{e} \geq 0$, which can be thought of as capacities. These regulate how passengers flow on the network, as higher conductivity is allocated to edges that are more utilized, while low-conductivity edges are those with fewer passengers. Hence, determining the values of $\mu_{e}$, $\forall e \in E$, implies determining the flows of passengers, and therefore of traffic on the network. The distribution of conductivities is regulated by the following dynamics and main equations of our model:
\begin{align}
\label{eqn:kirchoff}
\sum_{u} \Lap_{\Inode \Inodetwo} \Press_{\Inodetwo}^\Icomm &= \NetMass_\Inode^\Icomm \qquad\qquad\qquad  \forall v \in V,  \forall \Icomm = 1,\dots,\Ncomm \\
\label{eqn:dynamics}
\displaystyle\f{d \Cond_\Iedge}{dt} &= \Cond_\Iedge^{\beta-2} f (F_e)  - \Cond_\Iedge \qquad\forall e \in E.
\end{align}
Here $L$ is the weighted  Laplacian matrix of the network, with entries $\Lap_{\Inode \Inodetwo} := \sum_\Iedge \,  (\Cond_\Iedge / \ell_\Iedge) \Inc_{\Inodetwo \Iedge}\, \Inc_{\Inode \Iedge}$; $\Press_\Inode^\Icomm$ are pressure potentials generated by a commodity $\Icomm$ on the nodes; $f$ is a non-negative function of the fluxes $F_e$, $M$-dimensional vectors with entries $\Flux^\Icomm_\Iedge := \Cond_\Iedge (\Press^\Icomm_\Inodetwo - \Press^\Icomm_\Inode)/\ell_\Iedge$, for $e = (\Inodetwo, \Inode)$. A visualization of the main model's variables is shown in \cref{fig:model_representation}.

\Cref{eqn:kirchoff} is \kl~law, expressing conservation of mass; \cref{eqn:dynamics} regulates the time evolution of conductivities by means of a feedback mechanism where the higher the flux on an edge, the larger its conductivity $\mu_{e}$. All commodities share one unique infrastructure, so we follow \cite{lonardi2021designing} and assume that $\mu_{e}$ is the same for all $i$. This particular modeling choice corresponds to not prioritize any commodity in particular, i.e. having all users sharing the metro infrastructure without any hierarchy. However, one could consider imposing a set of rules for traffic regulation by explicitly accounting for different $\mu_e^i$ terms.

The growth of $\Cond_\Iedge$ is governed by the function $f$, that is typically an increasing and differentiable function of some norm of the fluxes \cite{bonifaci2021physarum, lonardi2021designing}. The aim of our work is to investigate how different expressions of $f$ result in different distributions of passengers flows, thus we focus on the following two choices: (i) $f(x) = ||x||_2^2$ ($2$-norm), and (ii) $f(x) = || x ||_1^2$ ($1$-norm). The first captures intuition in contexts as plant biology, where nutrients travel independently in conduits which are held together in fibers, contributing to growth of branches. However, it may not be the most appropriate one in applications involving transport of passengers, as the 2-norm does not have a straightforward interpretation. On the contrary, the latter is arguably a more natural choice, backed up by the intuition that edge capacities are controlled by the \emph{number} of passengers traveling on them (instead of the sum of squares). Both norms are taken squared, this is motivated by an analogy between our dynamics and Joule's law in electrical circuits, that we discuss in \cref{sec:connection_ot}.  We remark that other possible choices of $f$ can be used, e.g. the complete spectrum of $p$-norms,  or a tunable sigmoid profile as in \cite{tero2007mathematical}, these can be interesting subjects for future work. 
The effect of $f$ is balanced by a negative linear term in the conductivities, determining their exponential decay in time if no mass is moving through an edge. Note that our dynamics is highly non-linear in $\Cond_\Iedge$, since least-square solutions of \kl~law are of the form $\Press_\Inode^\Icomm = \sum_\Inodetwo \Lap^{\dagger}_{\Inode \Inodetwo} \NetMass_\Inodetwo^\Icomm$, with $\dagger$ denoting the Moore-Penrose inverse. Finally, the role of the free parameter $0 < \beta < 2$ is to capture different transportation mechanisms: $\beta > 1$ consolidates passengers on fewer edges, following a principle of economy of scale; $\beta < 1$ enforces passengers to distribute more broadly along the network; $\beta=1$ is shortest-path like.

\subsection{Connection with optimal transport}
\label{sec:connection_ot}

The dynamics introduced in \cref{sec:dyn} has a strong connection with optimal transport theory. In fact, in \cite{lonardi2021designing} it is shown that \emph{stationary solutions} of \cref{eqn:kirchoff,eqn:dynamics} are also stationary points of the minimization problem:
\begin{align}
\label{eqn:constrained_1}
\min_{\Flux \in \mathbb{R}^{|\Nedge| \times \Ncomm}} \; & J := \frac{1}{2} \sum_\Iedge \frac{\ell_\Iedge}{\Cond_\Iedge} f({\Flux_\Iedge}) \\
\label{eqn:constrained_2}
\text{s.t. \,\,}& \sum_{\Iedge} \ell_\Iedge \Cond_\Iedge^{2-\beta} = K^{2-\beta} \\
\label{eqn:constrained_3}
& \sum_\Iedge \Inc_{\Inode \Iedge} \Flux_\Iedge^\Icomm = \NetMass_\Inode^\Icomm \quad \forall \Inode \in V, \forall \Icomm=1,\dots,M,
\end{align}
for a fixed constant $K > 0$ and where $J$ is the dissipation cost. This is also equivalent to minimizing $J_{\Gamma} := \sum_\Iedge \ell_\Iedge f(\Flux_\Iedge)^\Gamma$, with $\Gamma = (2-\beta)/(3-\beta)$, generalizing Banavar \emph{et al.} \cite{maritan}.

The crucial distinction between the $1$-norm and $2$-norm dynamics is that the latter admits the Lyapunov function
\begin{align}
\label{eqn:lyapunov}
\mathcal{L}_\beta(\ccup{\mu_{e}}) := \frac{1}{2} \sum_{\Icomm,\Inode} \Press_\Inode^\Icomm \NetMass_\Inode^\Icomm + \frac{1}{2(2-\beta)} \sum_{\Iedge}  \ell_\Iedge \Cond_\Iedge^{2-\beta},
\end{align}
which enables to prove that \emph{asymptotics} of the dynamics minimize $J$ \cite{lonardi2021designing} for $\beta \leq 1$. We remark that for $\beta \leq 1$ the Lyapunov admits a unique minimum although possibly multiple minimizers, while for $\beta > 1$ the functional has several local minima. Noticeably, the first sum in \cref{eqn:lyapunov} is equivalent to $J=(1/2)\sum_e \ell_\Iedge ||\Flux_\Iedge||_2^2/ \Cond_\Iedge$ (see \cref{secAPX:lyap}).

The second term in \cref{eqn:lyapunov} is $W := (\sum_\Iedge \ell_\Iedge \Cond_\Iedge^\gamma ) / 2\gamma$ with $\gamma=2-\beta$, interpretable as the cost to build the network. With this in mind, the Lyapunov functional becomes the sum of dissipation and infrastructural costs.

As mentioned before theoretical guarantees cannot be recovered for the $1$-norm dynamics, where a Lyapunov functional is not straightforward to derive. While solving the dynamics may still result in meaningful flows, we cannot guarantee that these solutions minimize the cost $J_\Gamma$, i.e. to have optimal transport.

However, we find empirically that on the metro network of Paris---our case of study---$J_\Gamma$ decreases along solution trajectories of the dynamics, with stationary solutions lying in a basin of the cost. This empirical result is valid here, but this may not be true for other configurations of the network or initial conditions of the dynamics, hence practitioners should first validate their model (see \cref{secAPX:preprocessing,secAPX:validation} for a more detailed explanation; a listing of the variables introduced in \cref{sec:dyn,sec:connection_ot} can be found enclosed as Supplementary Information).

\subsection{Results on the Paris metro network}
\label{sec:results_paris}

In this work, we investigate the applicability of the dynamics in \cref{eqn:kirchoff,eqn:dynamics} on the Paris metro. Topology data are taken from \cite{kujala2018collection}, the network is preprocessed to have a total of $|V| = 302$ nodes and $|E| = 359$ edges, coherently with the observed metro of Paris (\cref{secAPX:preprocessing}). As anticipated, we define commodities as stations where passengers enter. This means that each vector $\NetMass^\Icomm$ has only one positive element in $\Inode = \Icomm$ (where the passengers of type $i$ enter), while the remaining elements of $\NetMass^\Icomm$ contain the outflows of passengers who travel from $\Inode$. Other choices can also be made based on the application, but this will not impact the validity of the model. Lastly, we introduce the parameter $0 \leq \rho \leq 1$. This averages the passenger inflows as $S_{v=i}^i (\rho)= S_{v=i}^i  - \rho ( S_{v=i}^i  - \langle S^i  \rangle )$, with $\langle \cdot \rangle$ average over the nodes. When $\rho$ tends to $1$ passengers distribute uniformly on the network, while $\rho$ approaching 0 means passengers enter and exit station more heterogeneously, see \cref{secAPX:validation}. 

\begin{figure*}[t]
\centering
\includegraphics[width=0.9\textwidth]{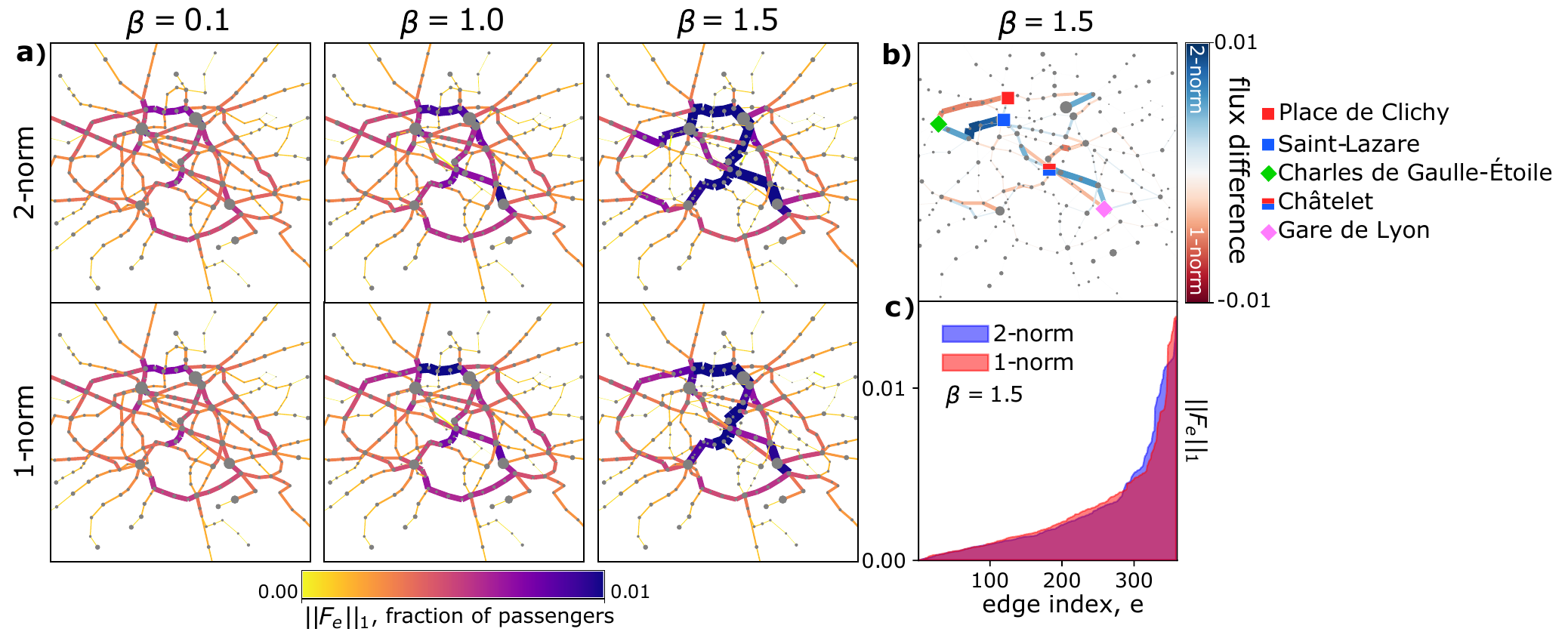}
\caption{\label{fig:panel_network} \textbf{Optimal transport networks panel.} \textbf{(a)} Optimal transport networks with $\beta = 0.1,1.0,1.5$ for the $1$-norm and $2$-norm. Edge thickness and color are proportional to $||\Flux_\Iedge||_1$, normalized to sum to $1$; node sizes are proportional to the number of passengers entering them. All the quantities are averaged over $100$ runs of the dynamics with $\Cond_\Iedge(0)\sim U(0,1)$. \textbf{(b)} Network colored using the difference of the fluxes obtained with the $1$-norm and with the $2$-norm. Results are displayed for $\beta = 1.5$, and using the data of \cref{fig:panel_network}a. Widths of edges are proportional to the absolute value of the flux difference, so that by matching the color and size information it is possible to distinguish  differences in networks generated by the two response functions. Marked stations are those discussed in \cref{sec:results_paris}. \textbf{(c)} Sorted flux distribution over the edges for $\beta = 1.5$. All quantities have been computed with $\rho = 0.0$, i.e. $S(\rho=0.0) = S$ (see \cref{secAPX:preprocessing,secAPX:validation}). Similar panels for $\rho=0.5$ and $\rho=1.0$ can be found in Supplementary Figs.  1 and 2.}
\end{figure*}

We test the two response functions $f$. Optimal fluxes resulting in the two cases can be seen in \cref{fig:panel_network}a, where the thickness of each edge is proportional to the fraction of passengers traveling through it. As expected, for $\beta < 1$ optimal transport networks are loopy, with many densely connected edges having fairly uniform fluxes. On the contrary, for $\beta > 1$ optimal topologies are more tree-like, with few central arteries where traffic is highly concentrated. This applies to both cases.

We notice two distinct behaviors, depending on $\beta$.
For $\beta < 1$ ($\beta = 0.1$ in \cref{fig:panel_network}a), solutions cannot be distinguished. This is explained by the Lyapunov functionals $\mathcal{L}_\beta$ being strictly convex in this case, with stationary conductivities that are their only minimum. This observation suggests that in the congested transportation regime ($0 < \beta <1$), where one aims at minimizing traffic congestions, using the 2-norm is equivalent to the more intuitive 1-norm formulation.
This is not the case for $\beta > 1$, where the two dynamics favor different local minima. These correspond to optimal networks with distinct central arteries where passengers are directed. The differences are further accentuated in \cref{fig:panel_network}b, where edges are colored with flux differences in these two cases, and where we highlight with markers instances of highly traversed stations. In detail, we can see that two routes branch from Charles de Gaulle-Étoile, the upper one passing by Place de Clichy is favored by the $1$-norm, and the lower one reaching Saint-Lazare is preferred by the $2$-norm. As for the connection between Châtelet and Gare de Lyon, we observe that the $1$-norm tends to favor the shortest path between the two stations, with most passengers travelling in a straight line. On the contrary, the path selected by the $2$-norm has a deflection.

\begin{figure}[t]
\centering
\includegraphics[width=0.5\textwidth]{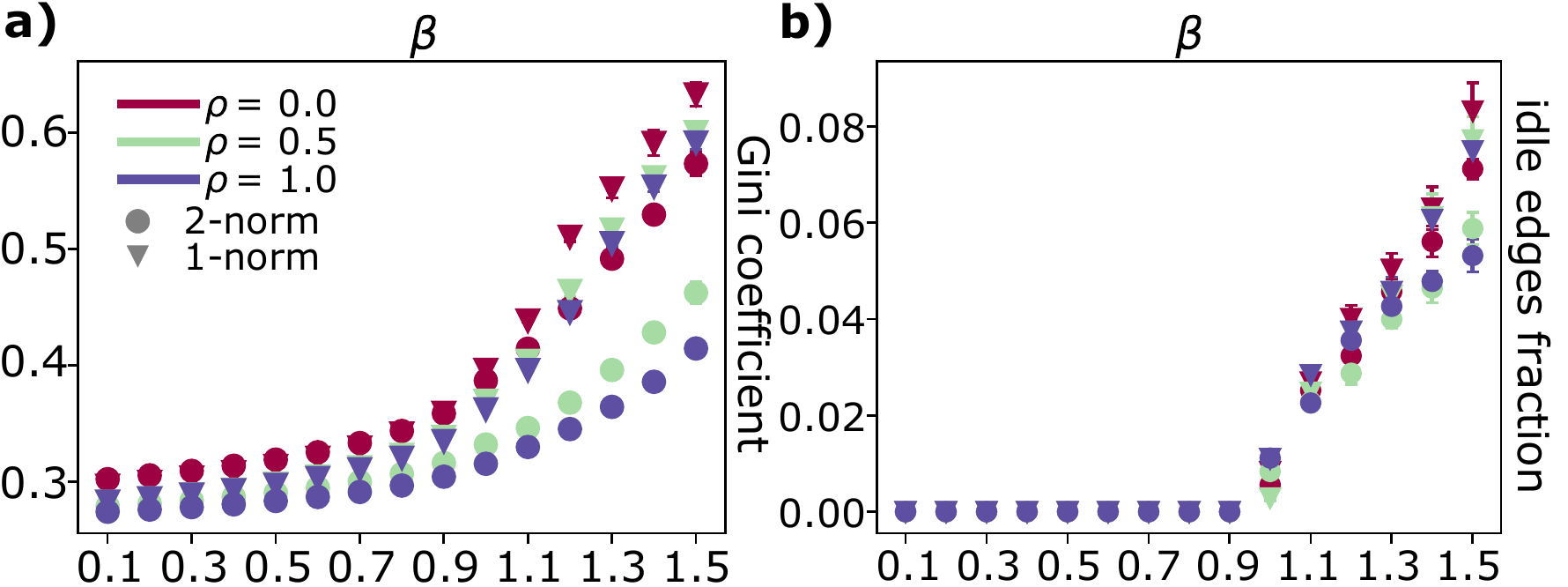}
\caption{\label{fig:idle_gini} \textbf{Evaluation metrics for the optimal transport networks}. \textbf{(a)} Gini coefficient vs. $\beta$. \textbf{(b)} Idle edges fraction vs. $\beta$. Each point is averaged over $100$ runs of the dynamics with random initializations of the conductivities.}
\end{figure}

Stimulated by these qualitative differences, we investigate different metrics for an in-depth quantitative evaluation for the case $\beta=1.5$. First, analyzing the sorted distributions of the fluxes $||\Flux_e||_1$ in \cref{fig:panel_network}c, we notice that the $1$-norm dynamics has a more pronounced fat-tailed distribution with a sharper and higher peak.
This means that the $1$-norm tends to concentrate fluxes on fewer edges. Such effect becomes starker for more homogeneous distributions of passengers entering the stations, i.e. setting $\rho = 0.5,1.0$ (see Supplementary Figs.  1 and 2).

We quantify this using the Gini coefficient \cite{dixon1987bootstrapping}:
\begin{align}
\text{Gini coefficient}(x) := \frac{1}{2 |\Nedge|^{2} \bar{x}} \sum_{m, n} \left| x_m-x_n \right|
\end{align}
for a quantity $x$, with $\bar{x} = \sum_\Iedge x_\Iedge / |\Nedge|$ being its mean, and $m,n$ denoting edges. In our analysis we set $x_e = ||F_e||_1$. Results are shown in \cref{fig:idle_gini}a, where the Gini coefficient is plotted against $\beta$ for different values of $\rho$.

As expected, the Gini coefficient increases with $\beta$, as users' paths are more concentrated along fewer edges. The values for the two dynamics are similar for $\beta < 1$, for the reasons mentioned above. Instead, for $\beta > 1$, markers progressively separate as $\beta$ increases. The $2$-norm has always smaller values than their counterparts, further demonstrating the tendency of the $2$-norm to dilute fluxes on a larger area of the network.

We study the behavior of the fraction of idle edges, i.e. the number of edges with negligible fluxes, divided by the total number of edges $|E|$, see \cref{fig:idle_gini}b. This quantity manifests a sudden phase transition at $\beta = 1$, where the dynamics switches from an homogeneous distribution of passengers on the entire network infrastructure, to a distribution progressively more concentrated on a smaller fraction of edges, as $\beta$ increases. Finally, the $2$-norm dynamics returns fewer idle edges than the $1$-norm, as paths are less concentrated. Notably, such abrupt phase transitions are typical of capacitated models on networks \cite{maritan}, and emerging in routing strategies involving a critical exponent regulating efficient transportation \cite{yan2006efficient}.

To summarize, we observe two main findings. First, we noticed that in the regime of $\beta < 1$ the $1$-norm and the $2$-norm produce identical optimal networks. This result does not hold for $\beta > 1$, where many local minima of $\mathcal{L}_\beta$ generate different optimal paths. Second, analyzing the fraction of idle edges, the Gini coefficient of the fluxes, and their distribution, we found that in the regime of branched transportation ($1<\beta<2$), the 2-norm tends to limit more traffic congestion, as paths are less consolidated into fewer edges compared to the 1-norm.

\subsection{Comparison with Dijkstra algorithm}

As discussed, the main property connecting our $2$-norm dynamics with optimal transport is that its stationary solutions are minimizers of the cost $J_\Gamma = \sum_\Iedge \ell_\Iedge || \Flux_\Iedge ||_2^{2\Gamma}$, with $\Gamma = (2-\beta)/(3-\beta)$ \cite{lonardi2021designing}. This cost, for $\beta = 1$ and $M=1$ is equivalent to that of \cite{bonifaci2012physarum, bonifaci2013short} and has optimal fluxes taking the shortest path from their source to their sink. A theoretical generalization of this result to the multicommodity setup is not trivial. In fact, for the $2$-norm case the cost reads $J_\Gamma = \sum_\Iedge \ell_\Iedge \sqrt{\sum_\Icomm (F_e^i)^{2}}$, that is not linear in the commodities, i.e. searching for its minimizer does not correspond to solving $M$ uni-commodity problems, one for each $\Icomm$, and then overlapping them. As for the $1$-norm, the dissipation cost with $\beta = 1$ is $J_\Gamma = \sum_\Iedge  \sum_\Icomm  \ell_\Iedge |F_e^i|$, and therefore its unique global minimum corresponds to that obtained overlapping $\Ncomm$ shortest paths.

We can numerically compare our methods with a shortest path routine using Dijkstra's algorithm \cite{dijkstra1959note}. Precisely, we iterate over the commodities and assign a flux $F_e^\Icomm$ equal to the fraction of passengers moving from the source $\Inode$ to the sink $\Inodetwo$, to each edge belonging the shortest path between $\Inode$ and $\Inodetwo$---the latter computed with Dijkstra's algorithm.

\begin{figure*}[htpb]
\centering
\includegraphics[width=0.85\textwidth]{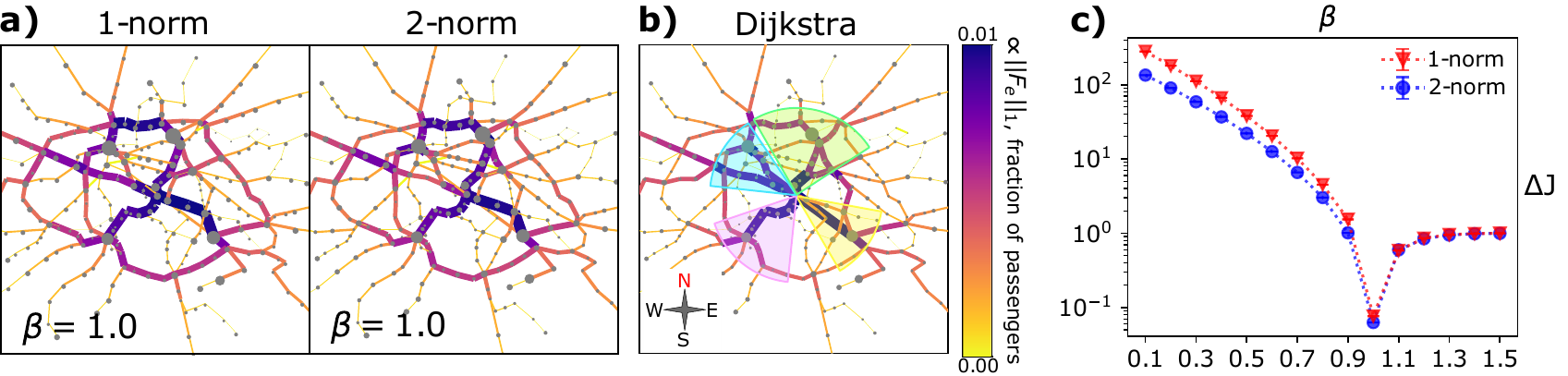}
\caption{\label{fig:dijkstra_panel} \textbf{Comparison between our methods and Dijkstra's algorithm.} Lenghts of edges are $\tilde{\ell}_\Iedge = \ell_\Iedge / n_\Iedge$, with $n_\Iedge$ number of vehicles passing through $\Iedge$ as in \cite{kujala2018collection}. This rescaling has been performed following the intuition that metro users moving along the network may travel using the fastest route (that for paths of the same length is the one with more frequent trains) to reach their destination; this may not correspond to the geographically shortest one. \textbf{(a)} Optimal transport networks for our methods. Quantities are computed over $100$ runs of the dynamics with random initialization of the conductivities, $\Cond_\Iedge \sim U(0,1)$. \textbf{(b)} Optimal transport network computed with Dijkstra's algorithm. For all three networks, edge widths and colors are $||\Flux_\Iedge||_1$, and the size of each node is proportional to the number of passengers entering in it. \textbf{(c)} Relative energy difference between our methods and Dijkstra's, taken in absolute value. Errorbars are standard deviations over $100$ realizations.}
\end{figure*}

We compare the optimal transport networks obtained using our methods with $\beta = 1$ (\cref{fig:dijkstra_panel}a) with the networks  returned by Dijkstra's algorithm (\cref{fig:dijkstra_panel}b).
 The three graphs are visibly similar but not identical. Particularly, we focus on the four highlighted areas in \cref{fig:dijkstra_panel}b, containing the main branches departing from the central area of the city of Paris. We see that the more trafficked routes in the pink South-West region are identical for our methods and for Dijkstra's one. Traffic in the Nort-West blue region seems to be more diluted for our methods, with the $2$-norm optimal network being slightly more similar to Dijkstra's. As for the North green region, both our algorithms concentrate traffic in a curved branch covering a large portion of the Northside of the city. This route is not prioritized in \cref{fig:dijkstra_panel}b, as traffic in the green portion is more distributed. Finally, in the South-East yellow area, there is only one main route branching from the city center, while its shape is straight for Dijkstra's, our methods favor a slight deflection.

We attribute these differences in the optimal topologies to the high complexity of the energy landscape of $J_\Gamma$. In fact, while Dijkstra's algorithm computes and overlaps each source-sink shortest path separately, our methods treat all the commodities at once. This may lead to convergence in suboptimal points, in particular around $\beta = 1$, where the cost transitions from being strictly convex ($\beta \leq 1$) to strongly non-convex ($\beta > 1$). While our method in this case may not always reach an optimal solution, it has the practical advantage of having a worst-case complexity of $O(M|V|^2)$ \cite{lonardi2021designing}. In principle, this can be further reduced using a backward Euler scheme combined with the inexact Newton-Raphson method for the discretization of \cref{eqn:dynamics}, and using a Multigrid solver for the solution of \cref{eqn:kirchoff} in $O(M|V|)$ steps \cite{facca2021fast}.  By contrast, Dijkstra's has a worst-case complexity of $O(|E|+|V|\log |V|)$ \cite{fredman1987fibonacci}, with the algorithms that needs to be executed $M^2$ (in our application to the Paris metro there are $M^2 = |V|^2$ source-destination pairs) times to solve a multicommodity problem.

Lastly, we test the deviation of the cost of our methods from Dijkstra's one. In \cref{fig:dijkstra_panel}c we plot the relative cost difference taken in absolute value, that is $\Delta J:= | J_\Gamma - J_\text{Dijkstra} | / J_\text{Dijkstra}$, with Dijkstra's network cost calculated as $J_\text{Dijkstra} = \sum_\Iedge \ell_\Iedge || \Flux_\Iedge ||_1$. This has a sharp drop at $\beta = 1$, where traffic is not favored nor penalized, with the cost of our network that is similar to the one of the shortest path returned by Dijkstra's algorithm. For $\beta < 1$ we have $J_\Gamma > J_\text{Dijkstra}$, showing that penalizing traffic congestion has the drawback of producing more expensive infrastructures. We observe the opposite behavior for $\beta > 1$, where $J_\Gamma < J_\text{Dijkstra}$, with congested networks that are progressively cheaper as $\beta$ increases.

\subsection{Network robustness to failures}
\label{sec:failures_sec}

We now showcase a possible relevant application of our model by analyzing network's robustness to structural failures as nodes removal.  Network managers interested in finding which stations are crucial for alleviating potential traffic overload can look at  the congested transportation regime (we set $\beta = 0.1$ to favor homogeneous fluxes) and investigate how fluxes resulting from our model distribute along the network.

\begin{figure*}[htpb]
\centering
\includegraphics[width=.95\textwidth]{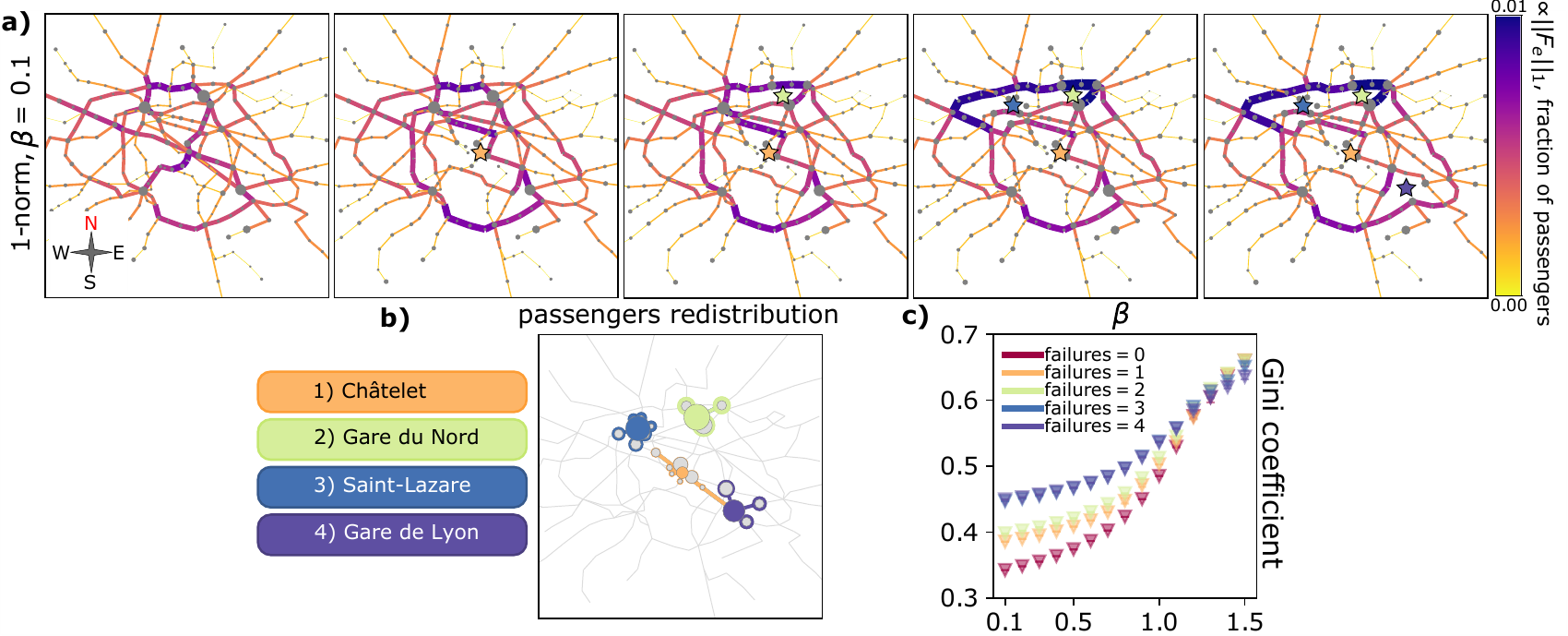}
\caption{\label{fig:broke_norm1} \textbf{Traffic rerouting after network structural failures.} \textbf{(a)} We plot the optimal transport networks after nodes trimming. Edge widths and color are normalized fluxes, the size of each node is proportional to the number passengers entering in it. All quantities are averaged over $100$ runs of the dynamics with random initialization of the conductivities, $\Cond_\Iedge \sim U(0,1)$. Stars highlight positions of removed stations, following the same scheme of \cref{fig:broke_norm1}b. \textbf{(b)} Network showing which stations have been removed, these correspond to the fully colored nodes, with colors chosen according to the legend on the left. Colored edges, and nodes with colored borders are those where the passengers get redirected. The colored borders are proportional to the passengers' growth. \textbf{(c)} Gini coefficients vs. $\beta$, errorbars are standard deviations. Points are colored following the scheme used in the rest of the panel. Similar results for the $2$-norm dynamics are in Supplementary Fig. 3.}
\end{figure*}

In detail, we remove sequentially a total of four stations from the network: Châtelet, Gare du Nord, Saint-Lazare, and Gare de Lyon. The last three are those with the largest number of inflowing passengers, while Châtelet has a central position and a high node degree $d=8$. Once each station was removed, its passengers were redirected to its neighboring nodes, and then solutions of the dynamics were found with this setting, as depicted in \cref{fig:broke_norm1}.

In \cref{fig:broke_norm1}a we display the $1+4$ networks obtained removing none, and the stations indicated in \cref{fig:broke_norm1}b. In \cref{fig:broke_norm1}c we plot the Gini coefficients of the optimal transport networks against $\beta$. We notice that for $\beta > 1$ all the points collapse together, regardless of the number of failures. This scenario, however, is of little interest for the situation we want to address, being flux aggregation already favored by $\beta > 1$.
As for $\beta < 1$, the difference in Gini coefficient gets wider the lower the $\beta$, with the largest gap at $\beta = 0.1$, we thus investigate this case.

Removing Châtelet from the network causes a considerable jump in the Gini coefficient, thus increasing the possibility of traffic jams. In fact, as we see from the second plot in \cref{fig:broke_norm1}a, all the passengers who were traveling on the South-West route branching from the city center are redirected in a way that congests southern arteries of the network. Removing Gare du Nord is not as crucial for traffic rerouting. Indeed, the main difference between the second and the third network of \cref{fig:broke_norm1}a is that passengers who were departing from Gare du Nord move to its southern neighboring station, Gare de l'Est, and modify only slightly their path. A large jump in the Gini coefficient is visible after removing Saint-Lazare, which seems to be fundamental for connecting the central area of Paris to its north side. In the fourth plot in \cref{fig:broke_norm1}a, we can see that traffic becomes highly congested in the northern branch directed from east to west. Gare de Lyon causes a negligible change in the Gini coefficient, associated to a modest traffic rerouting in the South-East part of the network.

\subsection{Pareto front}

\begin{figure}[htpb]
\centering
\includegraphics[width=0.3\textwidth]{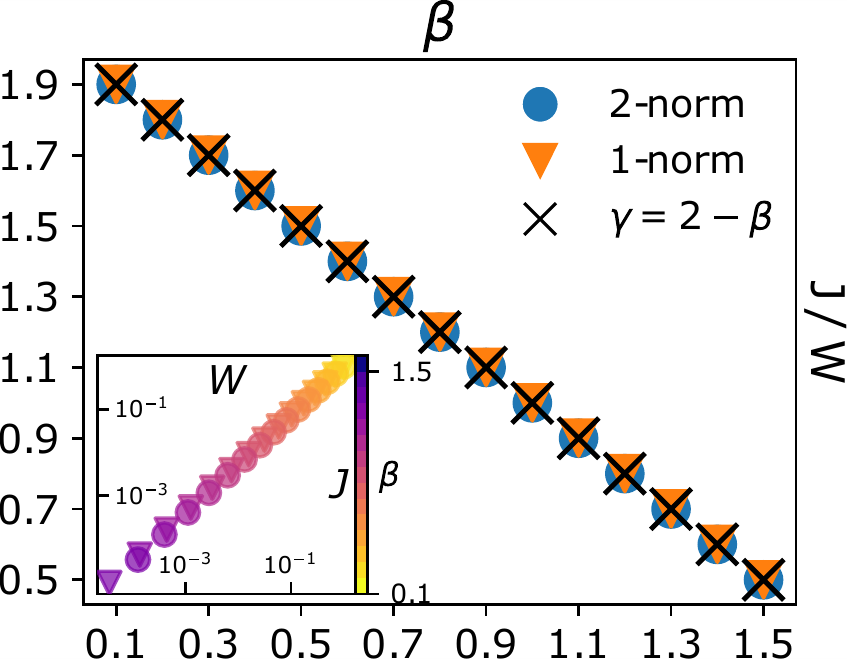}
\caption{\label{fig:pareto_scaling} \textbf{Pareto front.} We plot the dissipation/infrastructure cost ratio vs. $\beta$. Different points are averaged over $100$ runs with random initialization of the conductivities, $\Cond_\Iedge(0) \sim U (0,1)$. Inset: infrastructure cost, $W$, vs. dissipation cost, $J$. Marker shapes are identical to those of the main plot, colors follow the colorbar over $\beta$.}
\end{figure}

To conclude our analysis of the multicommodity routing problem it is possible to verify that stationary solutions of \cref{eqn:kirchoff,eqn:dynamics} lie in the Pareto front (\cref{fig:pareto_scaling}), which can be expressed in closed form as:
\begin{align}
\label{eqn:pareto_closed}
\frac{J}{W} = \gamma,
\end{align}
with $\gamma = 2 - \beta$ (see \cref{secAPX:pareto}).

The emergence of a Pareto front between $J$ and $W$ is not limited to engineering networks like the ones studied here. A similar trade-off has been observed in the widened pipe model for plants of Koçillari \cite{kocillari2021widened}, where minimization of hydraulic resistance and of carbon cost  compete for natural selection.

Moreover, looking at the inset of \cref{fig:pareto_scaling} and \cref{fig:idle_gini} we can observe that the Gini coefficient and the fraction of idle edges can be interpreted as driving forces responsible for the design of the optimal transport network, counterbalancing its cost. In fact, congested transport networks obtained for low values of beta $\beta$ have a high cost, but are more resilient to damage---low Gini and no idle edges---being their infrastructure densely connected.  On the contrary, setting $\beta$ large has the effect of producing sparse networks. These infrastructures have the benefit of being cheaper, but they are less resistant to node and edge failures, as mentioned in \cref{sec:failures_sec}.

\section{Conclusions}

Multicommodity routing is a powerful tool to model optimal network configurations in transportation systems \cite{salimifard2020multicommodity}. In this work, we developed a robust and efficient model able to perform this task by finding stationary solutions of a dynamical system controlling fluxes and conductivities of edges. Our dynamics extends previous works focusing solely on the unicommodity \cite{facca2019numerics,facca2020branching,facca2016towards,bonifaci2012physarum,hu2013adaptation}, and on the multicommodity setup \cite{adinoyi2021optimal,lonardi2021designing,bonifaci2021physarum}.

Precisely, we propose two different response functions regulating the growth of conductivities, whose evolution is dictated by the passengers moving in the metro. We performed a thorough empirical study of the optimal transport networks resulting in the two cases. Using metrics like the fraction of idle edges and the Gini coefficient of the fluxes, we found that the two functions behave similarly in the congested transportation regime, but differently in the branched transportation one. In this case, the 1-norm dynamics produces flows that are more concentrated on fewer edges, potentially leading to traffic overload.
We addressed the capability of our method to recover shortest path networks by comparing it with Dijkstra's routine. Such comparison showed that our approach is a viable computational alternative to perform this task, achieving accurate results and being, in principle, scalable for large networks. Additionally, we performed an experiment to measure network robustness to infrastructural failures, revealing that the stations of Châtelet and Saint-Lazare are crucial to ease congestion of metro routes. Finally, we showed that solutions of our model lie in the Pareto front drawn by the energy dissipated during transport and the network infrastructural cost.

Altogether, our findings extend the current research in multicommodity routing problems using optimal transport principles and help to understand the mechanism underlying passenger flows in transportation systems. 

Our formalism can be further extended to other possible applications related to the flow of passengers in transport networks. An example could be to incorporate time dependences in the passengers' inflows, modeling scenarios where stations are subject to different loads during a day, thus generalizing \cite{lonardi2021infrastructure}. One could also compare the extent of traffic jams in multicommodity settings to that of other routing strategies for urban transport displaying notable phase transitions and scaling laws \cite{toroczkai2004jamming, zhao2005onset, yan2006efficient}.

We would like to remark that our approach is applicable to a variety of practical problems unrelated to transportation systems. A practitioner may then consider response functions for the dynamics alternative to those studied in this work. The analysis performed in this work show how such a problem can be addressed and paves the way for further research beyond urban transportation networks.

\section{Methods}
\label[Methods]{sec:methods}

\subsection{Lyapunov and dissipation cost equivalence}
\label[Methods]{secAPX:lyap}

Here we show that the first term of the Lyapunov functional in \cref{eqn:lyapunov} is identical to the $2$-norm dissipation cost $J=(1/2)\sum_{e} \ell_{e} ||F_{e}||^{2}_{2} / \mu_{e}$, we follow \cite{lonardi2021designing}. Multiplying both sides of \cref{eqn:kirchoff} for $p_v^i$ and summing over $i$ and $v$ yields
\begin{align}
\label{eqn:derivation_1}
\sum_{\Icomm, \Inode, \Inodetwo, \Iedge} (\Cond_\Iedge / \ell_\Iedge) \Inc_{\Inodetwo \Iedge}\, \Inc_{\Inode \Iedge} p_\Inodetwo^i p_\Inode^i &= \sum_{i,v}p_v^i S_v^i \\
\label{eqn:derivation_2}
\sum_e \f{\ell_e}{\mu_e} ||F_e||_2^2 &= \sum_{i,v}p_v^i S_v^i,
\end{align}
where we made explicit the network Laplacian entries $\Lap_{\Inode \Inodetwo} := \sum_\Iedge \,  (\Cond_\Iedge / \ell_\Iedge) \Inc_{\Inodetwo \Iedge}\, \Inc_{\Inode \Iedge}$, and we used the definition of the fluxes $\Flux^\Icomm_\Iedge := \Cond_\Iedge (\Press^\Icomm_\Inodetwo - \Press^\Icomm_\Inode)/\ell_\Iedge$, for $e = (\Inodetwo, \Inode)$, and $\forall i$. \Cref{eqn:derivation_2} is the identity we wanted to prove.

\subsection{Preprocessing}
\label[Methods]{secAPX:preprocessing}

The original dataset in \citep{kujala2018collection} is provided as a multilayer network embedded with different transportation types, thus we performed a preprocessing to extract the metro network. First, we trimmed nodes belonging to other layers and then merged redundant stations having the same name by collapsing them together. This redundancy was due to the presence of stations with two entrances located in slightly different geographical positions; their coordinates displacement was always negligible compared to the physical extension of the whole network. The trimmed graph reflects consistently the real topology of the Paris metro. For convenience, the longitude and the latitude of nodes are rescaled within the range $[0,1]$.

We did not have access to the exact travel routes data, so we assigned the entries of $S$ based on the ``importance'' of each station. In fact the number of users validating their tickets when entering a station, the only data at disposal, is easier to track than the number of exiting users together with their entrance station. In practice, we assigned $N - 1$ positive ``influence factors'' to each station $\Icomm$, one for each node $\Inodetwo \neq \Icomm$ where the users entering in $\Icomm$ can potentially exit: $r_{\Inodetwo}^\Icomm = g^\Inodetwo / \sum_{w \neq \Icomm } g^w$, instead $r^{\Icomm}_{\Inode = \Icomm}=0$, where $g^\Inode$ is the amount of users entering the metro from $\Inode=\Icomm$.  Note that $0 \leq r_v^i \leq 1$ for all $v$ nodes, and $\sum_{v\neq \Icomm} r_v^\Icomm = 1$. Thus, we can estimate the number of people exiting from a station $\Inodetwo \neq \Icomm $ by assigning $\NetMass^\Icomm_\Inodetwo = - r^{\Icomm}_{u}\, g^{\Inode=\Icomm}$, while $\NetMass_{\Inode=\Icomm}^\Icomm = g^\Inode$. The intuition is that a station with a high entering volume of passengers, i.e. high $g^{\Inode}$, should have a large amount of exiting users, thus its ``influence'' value $r$ should be high.

\begin{figure}[t]
\centering
\includegraphics[width=.55\textwidth]{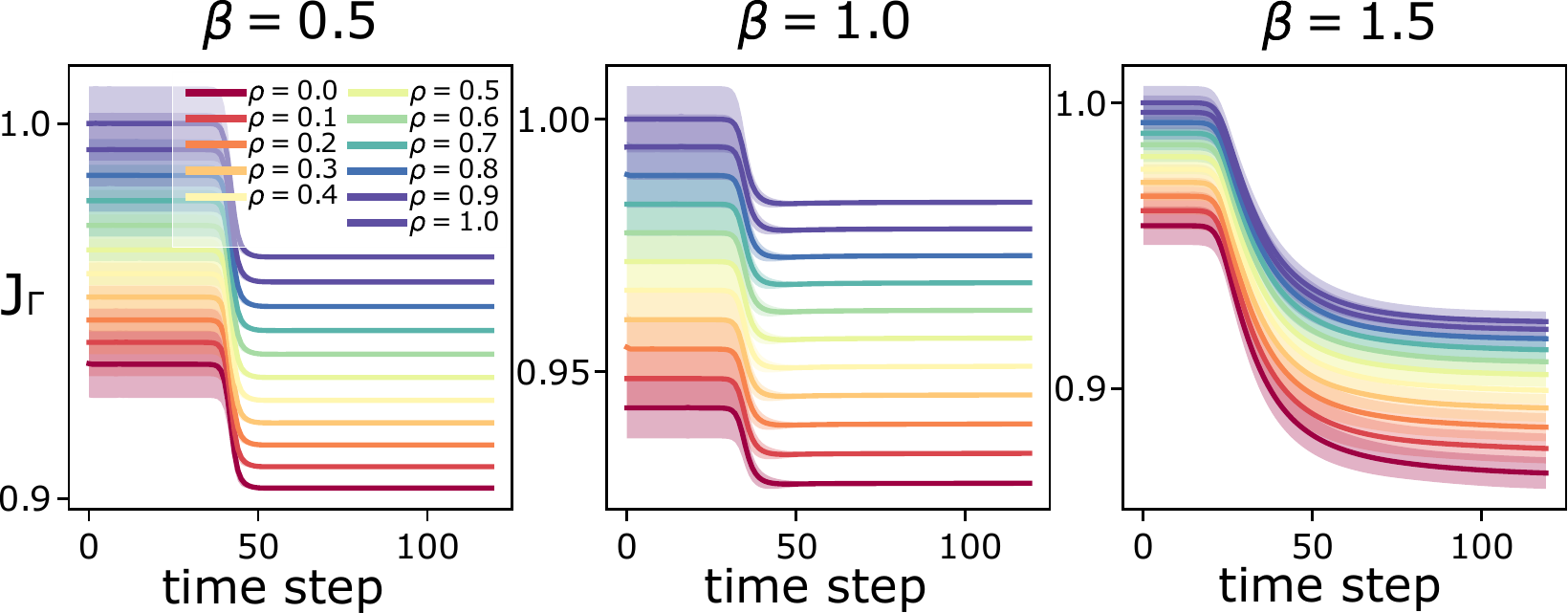}
\caption{\label{fig:cost_validation} \textbf{Validation of the dynamics with the $1$-norm. } We show the dissipation cost evolution along solution trajectories of \cref{eqn:kirchoff,eqn:dynamics}. Results are displayed for different combinations of $\rho$ and $\beta$, and are averaged over $100$ runs with random initializations of $\Cond_\Iedge(0) \sim U(0,1)$. The curves are normalized in $(0,1)$. Shaded areas denote standard deviations, these are thicker for $\beta > 1$ since the cost is concave, with a rich landscape a local minima.}
\end{figure}

\subsection{Validation}
\label[Methods]{secAPX:validation}

We validate \cref{eqn:kirchoff,eqn:dynamics} with the $1$-norm for several combinations of $\beta$ and of the input loads $\NetMass(\rho)$, with $0 \leq \rho \leq 1$ progressively smoothing the passengers inflows data collected in \cite{ratp2019}.  In detail,  users entering stations are regulated as $g^\Inode(\rho)= g^\Inode - \rho ( g^\Inode - \langle g \rangle )$ where $g^\Inode$ are inflows in $v$ as in \cite{ratp2019},  and $\langle \cdot \rangle$  averages over the nodes. Using this procedure, we build $S(\rho)$ following the ``influence assignment'' described in \cref{secAPX:preprocessing}. Thus, $S(\rho=0) = S$ corresponds to the originally extrapolated mass matrix, while $\NetMass_{\Inode=\Icomm}^\Icomm(\rho=1) = \langle g \rangle$ and $\NetMass_\Inodetwo^\Icomm(\rho=1) = - \langle g \rangle / (|V|-1)$ for all $\Inodetwo \neq \Inode=\Icomm$. Meaning that for $\rho=1$ passengers move with uniform rates from each---and to---all stations. In \cref{fig:cost_validation} we plot the time-evolution of $J_\Gamma$, which decreases over time.

\subsection{Pareto front derivation}
\label[Methods]{secAPX:pareto}

To obtain the Pareto form in closed form as in \cref{eqn:pareto_closed} it is sufficient to exploit the scaling $\Cond_\Iedge \sim (\Flux_\Iedge)^\delta$, $\delta = 3 - \beta$, valid for stationary solutions of the multicommodity dynamics \cite{lonardi2021designing}. In particular,  it is immediate to recover \cref{eqn:pareto_closed} by rewriting $J$ in \cref{eqn:constrained_1} as a function of the conductivities $\Cond_\Iedge$.

\section*{Data availability}
All data used for the experiments on the Paris metro network are publicly available \cite{ratp2019,kujala2018collection}.

\section*{Code availability}
An open-source implementation of the code is accessible at \href{https://github.com/aleable/McOpt}{\texttt{https://github.com/aleable/McOpt}}.

\section*{Author contributions}

All authors contributed to developing the models. A.L. and C.D.B. conceived the experiments, analyzed the results and reviewed the manuscript. A.L. conducted the experiments. All authors have read and agreed to the final version of the manuscript.

\section*{Competing financial interests}
The authors declare no competing interests.

\section*{Acknowledgments}
We thank Enrico Facca for helpful discussions and for insightful comments improving the manuscript. We acknowledge the help of Daniela Leite for data preprocessing. The authors thank the International Max Planck Research School for Intelligent Systems (IMPRS-IS) for supporting Alessandro Lonardi.

\bibliography{bibliography}

\mbox{}
\clearpage
\newpage

\setcounter{equation}{0}
\setcounter{figure}{0}
\setcounter{section}{0}
\setcounter{table}{0}
\setcounter{page}{1}
\makeatletter
\renewcommand{\theequation}{S\arabic{equation}}
\renewcommand{\thefigure}{S\arabic{figure}}
\renewcommand{\thetable}{S\arabic{table}}

\section*{\LARGE{Multicommodity routing optimization for engineering networks: Supplementary Information (SI)}}





\section{Listing of important variables}

In order to facilitate readability, in \Cref{tab:table_params} we list some of the main parameters of our model.

\begin{table}[htpb]
\resizebox{\textwidth}{!}{%
\setlength{\extrarowheight}{.2em}
\begin{tabular}{ c |  c  | c | c } \toprule
 Variable {\;\;} & {\;\;}  Dimension {\;\;} & Definition {\;\;} & Interpretation \\ \midrule
 $G(V,E)$  & --- &  --- &{\;\;} Network, $V = $ set of nodes, $E = $ set of edges\\
 $\ell = \{ \ell_e \}$ & $|E|$ & --- & {\;\;} Euclidian length of the edges\\
 $B = \{B_{ve}\}$ & $|V| \times |E|$ & --- & Signed incidence matrix of $G$\\
 $M$ & Scalar & --- & Number of commodities\\
 $S = \{ S_v^i \}$ & $|V| \times M$ & --- &{\;\;} Mass matrix containing in/outflows of passengers\\
 $\mu = \{ \mu_e \}$ & $|E|$ & --- &{\;\;}  Edge conductivities\\
 $L = \{ L_{uv} \}$ & $|V| \times |V| $ & {\;\;} $L_{uv} := \sum_e (\mu_e/\ell_e) B_{ue} B_{ve}$ {\;\;} &{\;\;} Weighted Laplacian matrix of $G$\\
 $\beta$ & Scalar &  --- & Regulatory parameter for traffic congestion \\  
  $p = \{ p_v \}$ & $|V|$ &  --- & Pressure potentials on nodes\\ 
  $F = \{ F_e^i \}$ & $|E| \times M$ &  $F_{e}^i :=  \mu_e(p_u^i - p_v^i) / \ell_e$ & Fluxes on edges generated by the commodities\\  
  $f(\cdot)$ & Scalar function &  $f(\cdot) = || \cdot ||_1^2$; $f(\cdot) = || \cdot ||_2^2$ & {\;\;} Response function for fluxes coupling\\
  $J$ &{\;\;}Scalar function{\;\;}&  $J := (1/2) \sum_\Iedge {\ell_\Iedge} f({\Flux_\Iedge}) / {\Cond_\Iedge}$ & {\;\;} Dissipation cost\\
  $W$ & Scalar function &  $W := (\sum_\Iedge \ell_\Iedge \Cond_\Iedge^\gamma ) / 2\gamma$ & {\;\;} Infrastructure cost\\
  $J_\Gamma$ & Scalar function &  $J_{\Gamma} := \sum_\Iedge \ell_\Iedge f(\Flux_\Iedge)^\Gamma$ & {\;\;} Dissipation cost, unconstrained minimization problem\\
  $\mathcal{L}_\beta$ & Scalar function &{\;\;}$\mathcal{L}_\beta := \frac{1}{2} \sum_{\Icomm,\Inode} \Press_\Inode^\Icomm \NetMass_\Inode^\Icomm + \frac{1}{2(2-\beta)} \sum_{\Iedge}  \ell_\Iedge \Cond_\Iedge^{2-\beta}${\;\;}& {\;\;} Lyapunov functional\\
  $\gamma, \delta, \Gamma$ & Scalars &{\;\;}$\gamma := 2-\beta$; $\delta:=1/(3-\beta)$; $\Gamma:=(2-\beta)/(3-\beta)${\;\;}& {\;\;} Auxiliary critical exponents\\
    $\rho$ & Scalar &  --- & {\;\;} Parameter for mass matrix smoothing\\ \bottomrule
\end{tabular}
}
\caption{\label{tab:table_params} Comprehensive listing of the main parameters and variables used.}
\end{table}

\section{Optimal transport networks}

\begin{center}
\begin{figure}[htpb]
\centering
\includegraphics[width=0.8\textwidth]{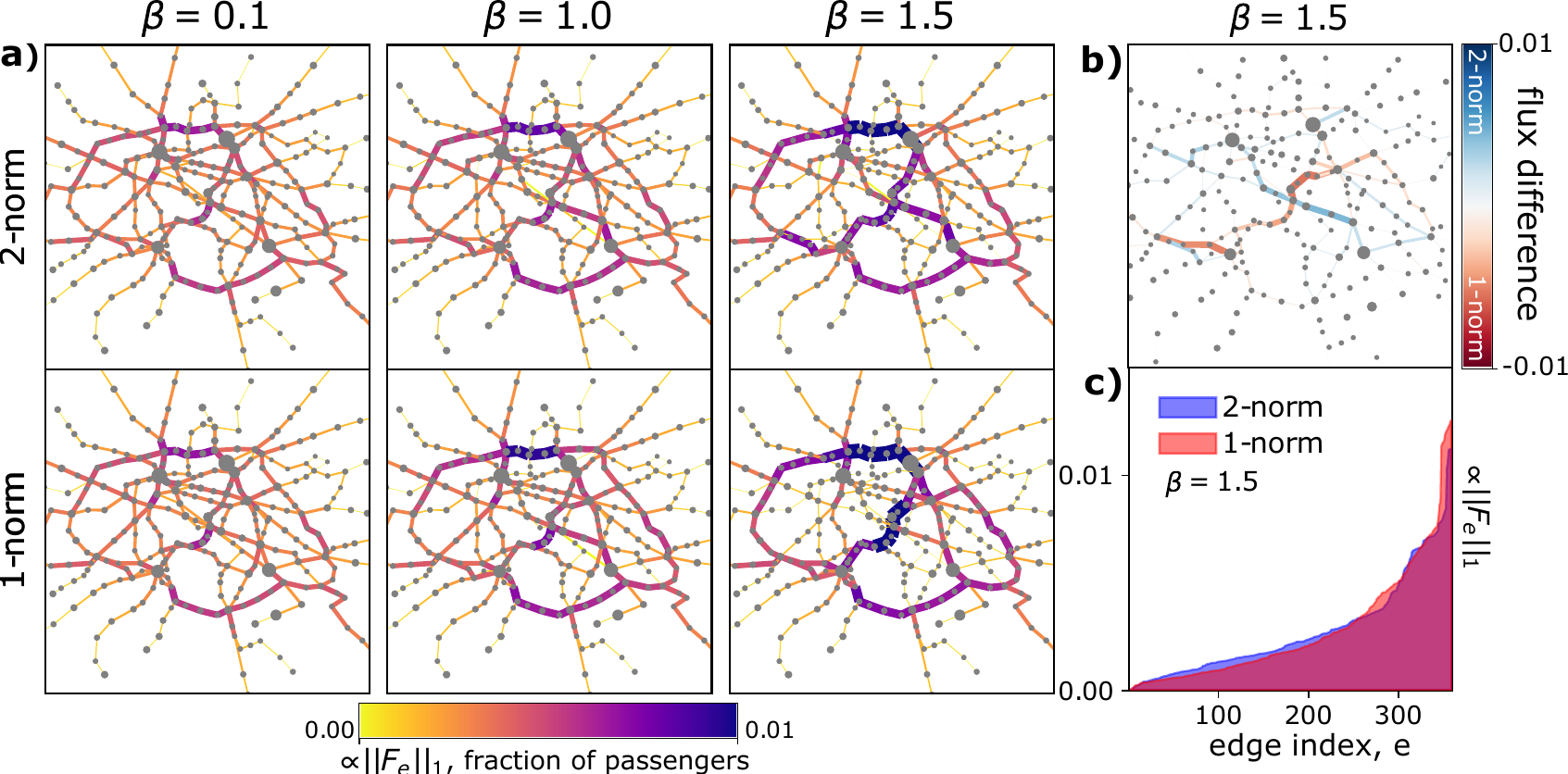}
\caption{\label{fig:panel_network_1_supp}\textbf{Optimal transport networks panel with forcing $S(\rho=0.5)$.} For a detailed description of the subplots one can refer to Fig. 2 in the main text.}
\end{figure}
\end{center}

\begin{center}
\begin{figure}[htpb]
\centering
\includegraphics[width=0.8\textwidth]{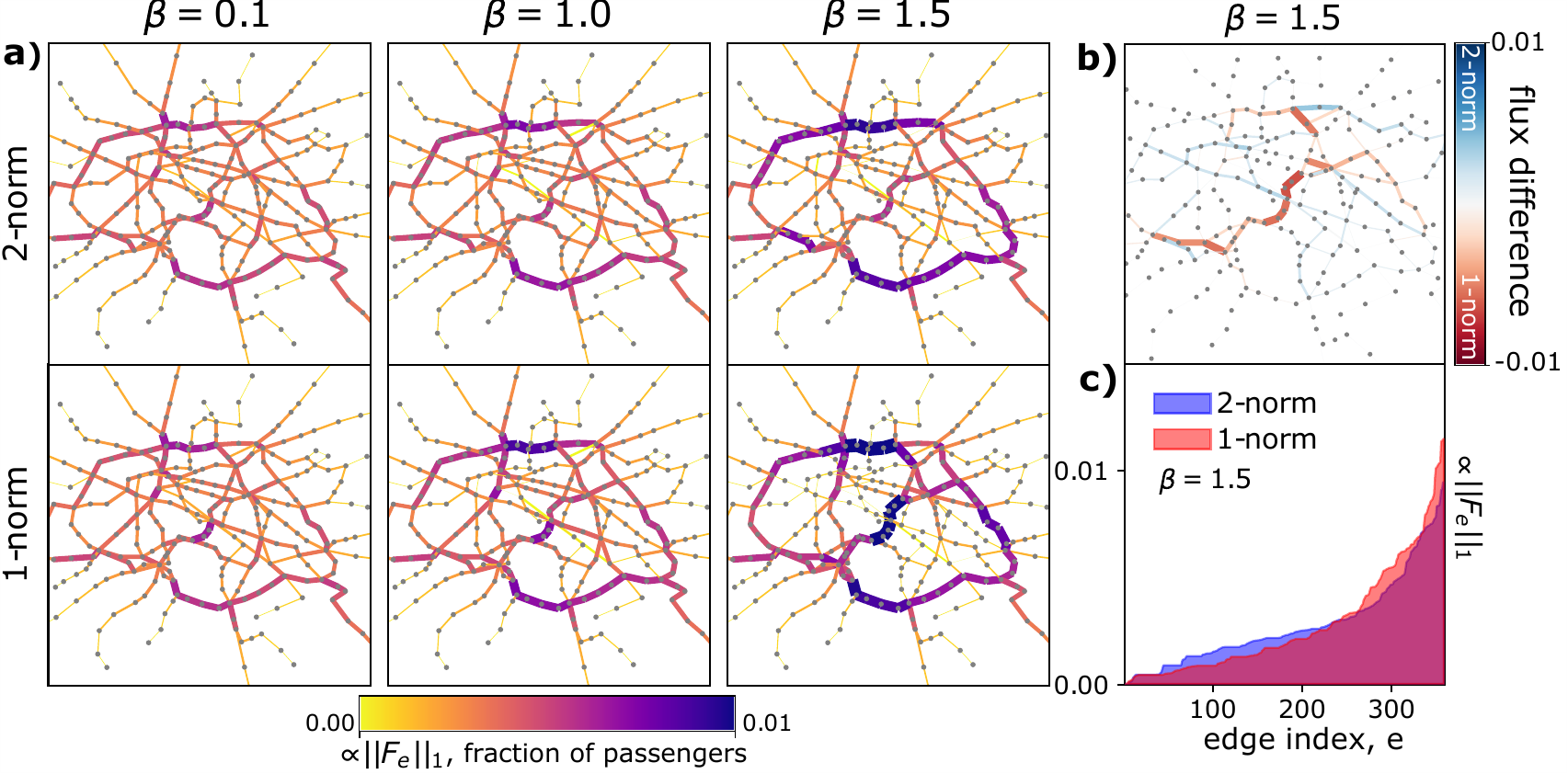}
\caption{\label{fig:panel_network_2_supp} \textbf{Optimal transport networks panel with forcing $S(\rho=1.0)$.} For a detailed description of the subplots one can refer to Fig. 2 in the main text.}
\end{figure}
\end{center}

In the panels in \cref{fig:panel_network_1_supp} and \cref{fig:panel_network_2_supp} we show the optimal transport networks for different configuration of the input forcings $S(\rho)$. In detail, we display the results for $\rho = 0.5$ (\cref{fig:panel_network_1_supp}), and those for $\rho = 1.0$ (\cref{fig:panel_network_2_supp}). Looking at the rightmost networks ($\beta=1.5$) of \cref{fig:panel_network_1_supp}a and \cref{fig:panel_network_2_supp}a one can observe a clear tendency of the $1$-norm dynamics to concentrate traffic more than the $2$-norm one. This trend reflects on the sorted distributions plotted in \cref{fig:panel_network_1_supp}c and \cref{fig:panel_network_2_supp}c, where the fluxes are more fat-tailed and homogeneous for the $2$-norm. Notably, the effect becomes starker the more the input inflows of passengers distribute uniformly on the nodes, i.e. increasing $\rho$.

\section{Network resistance to failures: 2-norm dynamics}

\begin{center}
\begin{figure}[!h]
\centering
\includegraphics[width=.85\textwidth]{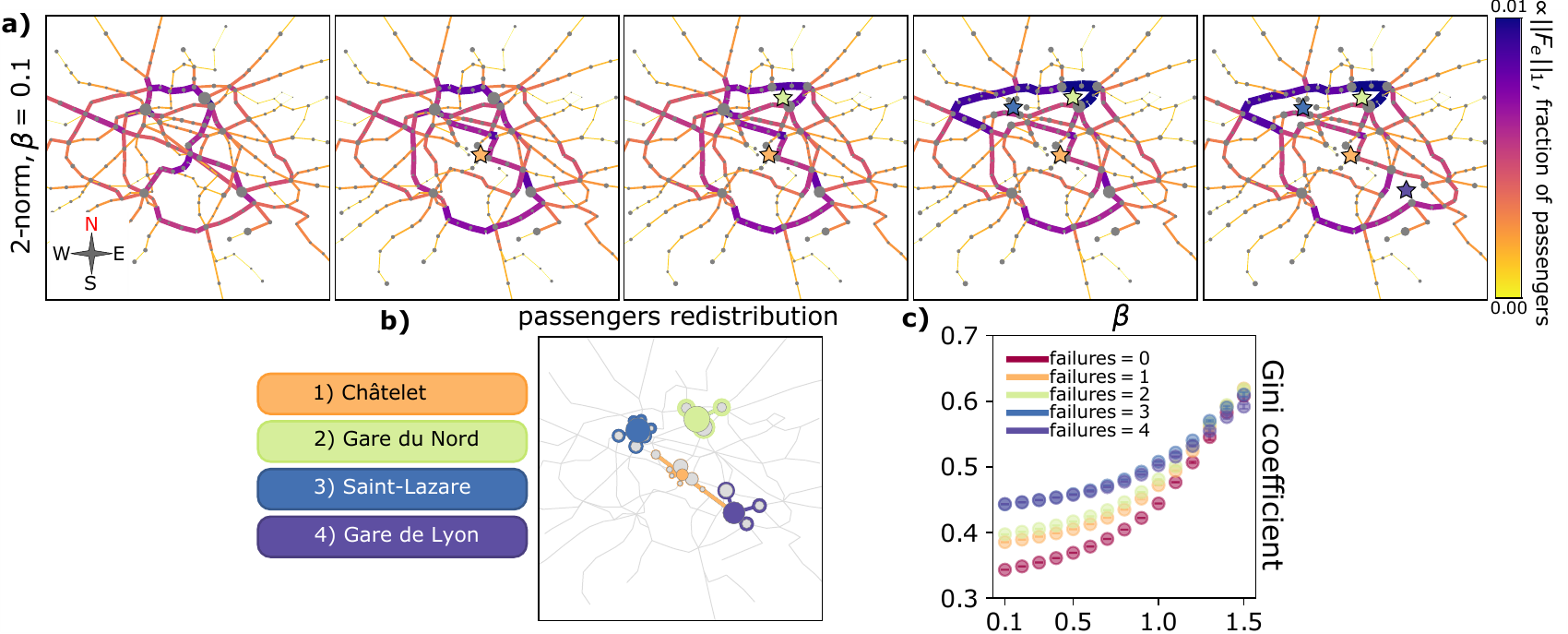} \caption{\label{fig:broke_norm2}\textbf{Traffic rerouting after network structural failures ($2$-norm dynamics).} For a detailed description of the subplots one can refer to Fig. 5 in the main text.}
\end{figure}
\end{center}

In \cref{fig:broke_norm2} we reproduce the experiments designed to test the resilience of optimal networks to node failures. Overall, results are similar to those in the main text, it is worth mentioning how the Gini coefficient values \cref{fig:broke_norm2}c are higher than the correspondent ones for the $1$-norm, symptom of the tendency of the latter forcing function to aggregate traffic. Another implication of this effect is that the Gini coefficient values for the $2$-norm tend to separate more for higher $\beta$ than those of the $1$-norm. In fact, the latter tend to be overlapped, regardless of the number of failures, on a larger portion of the $x$-axis (where $\beta > 1$).

\end{document}